\documentclass[12pt,preprint]{aastex}
\usepackage[]{natbib}

\begin{document}

\def\wisk#1{\ifmmode{#1}\else{$#1$}\fi}

\def\lt     {\wisk{<}}
\def\gt     {\wisk{>}}
\def\le     {\wisk{_<\atop^=}}
\def\ge     {\wisk{_>\atop^=}}
\def\lsim   {\wisk{_<\atop^{\sim}}}
\def\gsim   {\wisk{_>\atop^{\sim}}}
\def\kms    {\wisk{{\rm ~km~s^{-1}}}}
\def\Lsun   {\wisk{{\rm L_\odot}}}
\def\Zsun   {\wisk{{\rm Z_\odot}}}
\def\Msun   {\wisk{{\rm M_\odot}}}
\def\um     {$\mu$m}
\def\mic     {\mu{\rm m}}
\def\sig    {\wisk{\sigma}}
\def\etal   {{\sl et~al.\ }}
\def\eg     {{\it e.g.\ }}
 \def\ie     {{\it i.e.\ }}
\def\bsl    {\wisk{\backslash}}
\def\by     {\wisk{\times}}
\def\half {\wisk{\frac{1}{2}}}
\def\third {\wisk{\frac{1}{3}}}
\def\nwm2sr {\wisk{\rm nW/m^2/sr\ }}
\def\nw2m4sr {\wisk{\rm nW^2/m^4/sr\ }}

\title{Measuring the dark flow with public X-ray cluster data}

\author{A. Kashlinsky\altaffilmark{1}, F.  Atrio-Barandela\altaffilmark{2},
 H.  Ebeling\altaffilmark{3}
\altaffiltext{1}{SSAI and Observational Cosmology Laboratory, Code
665, Goddard Space Flight Center, Greenbelt MD 20771;
alexander.kashlinsky@nasa.gov} \altaffiltext{2}{Fisica Teorica,
  University of Salamanca, 37008 Salamanca, Spain}
   \altaffiltext{3}{Institute for Astronomy, University of Hawaii, 2680
  Woodlawn Drive, Honolulu, HI 96822}
}

\begin{abstract}
We present new results on the "dark flow" from a measurement of the dipole in the distribution of peculiar velocities of galaxy clusters, applying the methodology proposed and developed by us earlier. Our latest measurement is conducted using new, low-noise 7-yr WMAP data as well as an all-sky sample of X-ray selected galaxy clusters compiled exclusively from published catalogs. Our analysis of the CMB signature of the kinematic Sunyaev-Zeldovich (SZ) effect finds a statistically significant dipole at the location of galaxy clusters. The residual dipole outside the cluster regions is small, rendering our overall measurement 3-4 sigma significant. The amplitude of the dipole correlates with cluster properties, being larger for the most X-ray luminous clusters, as required if the signal is produced by the SZ effect. Since it is measured at zero monopole, the dipole can not be due to the thermal SZ effect. Our results are consistent with those obtained earlier by us from 5-yr WMAP data and using a proprietary cluster catalog. In addition, they are robust to quadrupole removal, demonstrating that quadrupole leakage contributes negligibly to the signal. The lower noise of the 7-yr WMAP also allows us, for the first time, to obtain tentative empirical confirmation of our earlier conjecture that the adopted filtering alters the sign of the KSZ effect for realistic clusters and thus of the deduced direction of the flow. The latter is consistent with our earlier measurement in both the amplitude and direction. Assuming the filtering indeed alters the sign of the KSZ effect from the clusters, the direction agrees well also with the results of independent work using galaxies as tracers at lower distances.
We make all maps and cluster templates derived by us from public data available to the scientific community to allow independent tests of our method and findings.
\end{abstract}

\section{Introduction: early peculiar-velocity measurements and ``dark flow"}

Peculiar velocities play an important role in understanding the large-scale gravitational field in the Universe and have been the subject of intense investigations over the past decades. Early determinations of peculiar velocities were based on surveys of
individual galaxies (see review by Strauss \& Willick 1995). First measurements by
Rubin and co-workers found peculiar flows of ${\sim}700$ km
s$^{-1}$ (Rubin et al 1976), but were largely dismissed at the time. A group collectively known as the
``Seven Samurai" found that elliptical galaxies within
${\sim}60h^{-1}$Mpc were streaming at ${\sim}600$ km s$^{-1}$ with
respect to the CMB (Dressler et al 1987, Lynden-Bell et al 1988). Using mainly spiral galaxies, Mathewson et al (1992) found that this flow does not converge until scales
much larger than ${\sim}60 h^{-1}$ Mpc, in agreement with the
results of a later analysis by Willick  (1999). With brightest cluster
galaxies as distance indicators for a sample of 119 rich clusters,
Lauer \& Postman (1994 - LP) measured a bulk flow of ${\sim}$700 km
s$^{-1}$ on a scale of ${\sim}150h^{-1}$Mpc. An improved re-analysis
of these data by Hudson \& Ebeling (1997), however, found a reduced bulk
flow pointing in a different direction. Using early-type galaxies in 56 clusters, Hudson et al (1999) found a similar bulk flow as LP and on a comparable scale, but
again in a different direction. By contrast, a sample of 24 SNIa
by Riess et al (1997) showed no evidence of significant bulk
flows out to ${\sim}100 h^{-1}$ Mpc, and a similar conclusion was
reached in a study of spiral galaxies by Courteau  et al (2004).

A complementary technique aimed at constraining bulk motions reconstructs directly the peculiar gravity of the observed galaxy distribution and uses
measurements of the dipole in the distributions of light and matter.
The
dipole derived from the distribution of galaxies mapped in optical
surveys is nearly aligned with the one obtained for
infra-red-selected galaxies, but both are misaligned with respect
to the CMB dipole generated by the motion of our Local Group
relative to the CMB rest frame (Rowan-Robinson et al 2000), although this misalignment becomes less troublesome if one relaxes the light-tracing-mass assumptions (see discussion by Gunn 1988). Kocevski et al (2004) and Kocevski \&
Ebeling (2006) measured the dipole anisotropy of an all-sky
sample of X-ray-selected galaxy clusters to probe mass concentrations beyond the Great Attractor   and found that most of the peculiar velocity of the Local Group is due to overdensities at $\ga 150h^{-1}$Mpc.


All galaxy techniques lose sensitivity at distances approaching and greater than $\sim$50-100 Mpc. The Sunyaev-Zel'dovich (SZ) effect, produced by hot gas in galaxy
clusters, is uniquely suited to probe flows to larger
distances; moreover, it is independent of
redshift and not subject to the systematics plaguing studies
using empirical distance indicators. The kinematic part of the SZ effect (KSZ) is directly proportional to the cluster velocity with respect to the cosmic micorwave background (CMB).
Because of its smallness, the KSZ effect has, however, not yet been measured for individual
clusters; observations of six clusters at a wide range of redshifts out to $z\simeq 0.82$ yielded an upper limit of $V\la 1,500$ km s$^{-1}$ on a poorly defined scale (Benson et al 2003). Kashlinsky \& Atrio-Barandela (2000, hereafter KA-B) have proposed a method to measure large-scale flows using all-sky cluster catalog and CMB all-sky data, such as obtained with WMAP. KA-B identified a statistic (the dipole of the CMB temperature field evaluated at cluster positions) which preserves the KSZ component while integrating down other (noise) terms. However, the method requires a CMB filter that removes the  primary CMB (which is strongly spatially correlated) without significantly attenuating the KSZ bulk flow contribution; clearly not every filter will achieve this.

Kashlinsky et al (2008, 2009 - KABKE1,2) have applied the KA-B method to a large cluster catalog finding a surprising flow (dubbed the "dark flow") extending to at least 300$h^{-1}$Mpc. Following this, an independent study of Watkins et al (2010) combined the available galaxy data suppressing the sampling noise in the various surveys and showed that all data (with
the exception of the LP sample) agreed with a
substantial motion on a scale of ${\simeq}50-100 h^{-1}$Mpc with amplitude and direction in good agreement with the KABKE measurements. In a follow-up study, Kashlinsky et al (2010, KAEEK) revise  the statistical analysis of their original study\footnote{Keisler (2009) together with KAEEK and AKEKE pointed out that KABKE did not account for correlations of the residual primary CMB fluctuations in the 8 DA channels of the WMAP data used in their analysis.} and use a much expanded cluster catalog, binned by cluster X-ray luminosity ($L_X$) to demonstrate that the CMB dipole increases with the $L_X$-threshold as required by the KSZ origin of the signal; such an $L_X$ dependence of the dipole is inconsistent with it originating from some putative systematic effect from primary CMB fluctuations. KAEEK find that the "dark flow" flow extends to at least $\ga 800$ Mpc, twice the distance reported by KABKE. Atrio-Barandela et al (2010, AKEKE), developed a formalism to understand -- both analytically and numerically -- the uncertainties in measurements using the KABKE filter; the same formalism is applicable to any filtering scheme. In addition, AKEKE demonstrate that the KABKE filter removes primary CMB fluctuations down to the fundamental limit of cosmic variance, rendering it optimal for such studies.

Very recently, the dark flow results of KABKE/KAEEK have found support by a study (Ma et al 2010) using a compilation of galaxy distance indicators which reports the same "tilt" velocity as the dark flow and pointing in the same direction, within the calibration uncertainties discussed in KABKE2/KAEEK/AKEKE. On the other hand, the KABKE results have been challenged by Keisler (2009). Replicating the analysis of KABKE1,2 using a cluster catalog compiled from publicly available data, Keisler confirmed the central dipole values measured by KABKE2, but claimed that it has only marginal statistical significance. AKEKE (Sec.~4 and Fig.~5) have since shown that Keisler's error estimates are erroneous\footnote{Clearly correlations between 8 DA channels used in the studies can at most increase the KABKE1,2 errors by $\sqrt{8}$, whereas Keisler claimed a $\ga\sqrt{20}$ increase.} and largely due to him not having removed the monopole and dipole from the CMB maps {\it outside} the mask. In a more recent challenge Osborne et al.\ (2010) have likewise used publicly available X-ray cluster data, applied alternative filtering schemes and claimed not to be able to replicate the "dark flow" results.


\section{Cluster bulk-flow measurements with public cluster data}

In this study we construct the cluster catalog from public data available to Keisler (2009) and Osborne et al.\ (2010) and demonstrate that, with the filtering scheme developed by us earlier, application of the KA-B method yields a statistically significant CMB dipole which is perfectly consistent with the KAEEK results\footnote{The CMB maps and cluster masks on which this study is based can be obtained from \url{http://www.kashlinsky.info/bulkflows/data\_public}}. We further address the calibration uncertainties in such a measurement to demonstrate that the data very likely require a sign change of the KSZ term from filtering as explained in KAEEK, an uncertainty we hope to eliminate with the proper calibration of our future catalog and our planned application to the upcoming Planck data as was proposed by us earlier\footnote{\vspace{-1.5mm}{http://www.rssd.esa.int/SA/PLANCK/docs/Bluebook-ESA-SCI(2005)1\_V2.pdf}}.

\subsection{The filtering scheme for the KA-B method}

The CMB temperature field in the presence of a bulk flow can be written as:
$\delta = n+ \delta_{\rm CMB} + \delta_{\rm TSZ} + A_{\rm KSZ} \cos\theta$,
where $n$ is the instrument noise, and the last term represents the contribution
to the dipole caused by the KSZ signal from any bulk flow. We want to measure the
KSZ amplitude, $A_{\rm KSZ}\propto \tau V$, whose value is in general
quite small compared to the TSZ and (primary) CMB terms. KA-B
suggested to boost the weight of the KSZ term by measuring the
dipole of the CMB maps at all-sky cluster positions.

Because the primary CMB is spatially highly correlated, a filter needs to be designed
that removes this component without significantly
attenuating $A_{\rm KSZ}$; clearly not every filter will achieve this. KABKE1,2 defined a filter, described in detail in KABKE2 and AKEKE,
which belongs to a Wiener variety and removes the primary CMB fluctuations from the concordance $\Lambda$CDM model by minimizing the mean squared deviation of the CMB measurements
from noise, $\langle (\delta_{\rm CMB}-n)^2\rangle$. In multipole $\ell$-space it is given by $F_\ell = (C_\ell - C_\ell^{\rm \Lambda CDM})/C_\ell$, where $C_\ell, C_\ell^{\rm \Lambda CDM}$ are the power spectra of the CMB map and the theoretical model convolved with the beam, respectively. AKEKE develop a formalism to quantify the errors in the resultant dipole determination that can be applied to {\it any} filtering scheme and show that the KABKE filter removes the primary CMB fluctuations down to the fundamental limit of the cosmic variance.

The filtering does, however, not remove the TSZ component and is thus by itself
insufficient to isolate the KSZ contribution. Atrio-Barandela et al (2008, herafter AKKE) helped critically to overcome this issue by demonstrating explictily, with WMAP data, that
clusters are well described by the NFW profile (Navarro et al 1996) and that their X-ray temperature,
$T_X$, should then decrease toward cluster outer parts. KABKE1,2 used this
property, after further empirical tests, to suppress the TSZ contribution
by measuring the CMB dipole at cluster positions over larger apertures,
evaluating the final dipole at {\it zero monopole}. The latter
also insures that the TSZ contribution to the dipole measured at
the final aperture is small.

In order to isolate the KSZ dipole  KABKE and KAEEK, as well as this study, proceed as follows: 1) we first filter separately each of the foreground-subtracted WMAP maps with a filter that removes the primary CMB fluctuations from $\Lambda$CDM, 2) we identify, in the filtered maps and for each cluster configuration, the aperture where the monopole over the cluster pixels vanishes (along with the TSZ contribution to the dipole), 3) we establish, by measuring the dipole separately for clusters binned by  $z$ and $L_X$, that the signal originates from the KSZ term. The last step is particularly important to test for the presence of systematic effects since these are not likely to correlate with either the redshift or the X-ray luminosity of the clusters in the sample.

\subsection{Filtered CMB maps and their uncertainties}

The original foreground-reduced CMB maps had their monopole and dipole subtracted outside the Galactic mask. The maps are pixelized in the HEALPix format with $N_{\rm side}=512$ (Gorski et al 2005). An extra step in our pipeline described in KAEEK is the additional subtraction of the quadrupole from the maps outside the Galactic mask prior to filtering. This probes the possible leakage of a quadrupole signal due to masking effects, and also removes any relativistic contribution from the local velocity, $v$, down to $O[(v/c)^3]$ terms of the octupole.

\begin{figure}
\plotone{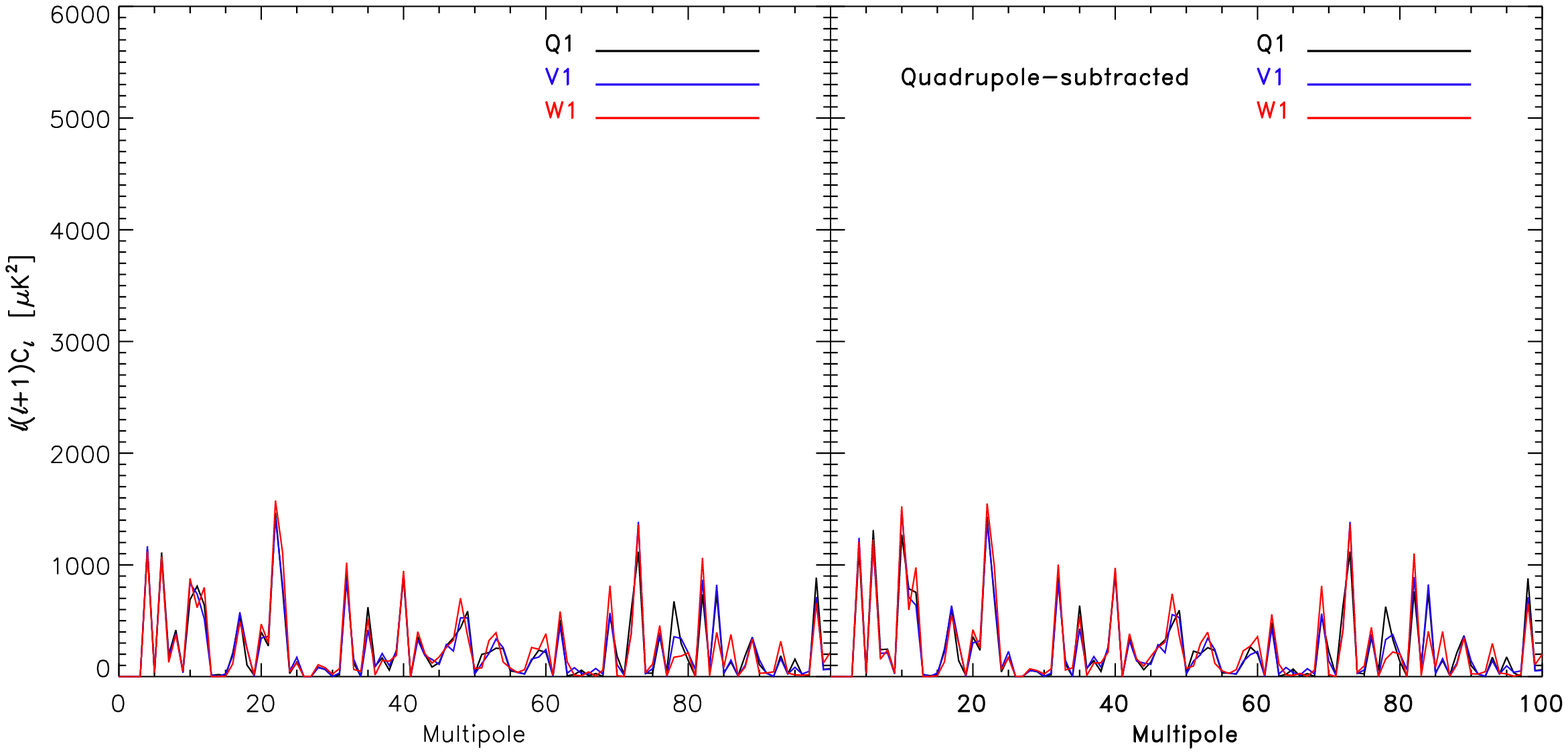} \caption{ \small{Spatial spectrum of the KABKE-filtered 7-year WMAP maps.  {\bf Left}: monopole and dipole subtracted maps. {\bf Right} Monopole, dipole {\it and} quadrupole-subtracted maps. }} \label{fig:f1}
\end{figure}

Fig. \ref{fig:f1} illustrates the properties of the filtered maps. There is no particular structure in $\ell$-space over any range of multipoles up to the scales subtending cluster apertures ($\sim 1^\circ$ radius).

To understand the measurability of the KSZ dipole from {\it filtered} 7-yr WMAP data it is instructive to estimate the uncertainties expected in a given filtering scheme. AKEKE developed analytical and numerical formalism which we briefly revisit below. The standard deviation of {\it any} filtered map is given by eq. 3 of AKEKE:
\begin{equation}
\sigma_{\rm map}^2 = \frac{1}{4\pi} \sum (2\ell+1) F_\ell^2 C_\ell
\label{eq:sigma_fil}
\end{equation}
Here $C_\ell = C_\ell^{\Lambda CDM} + N_\ell$ is the power spectrum of the original map containing the $\Lambda$CDM primary signal (convolved with the beam) and instrument noise, $N_\ell$. The uncertainty in measuring the monopole and three dipole terms from such maps is then $\simeq \sigma_{\rm map} \sqrt{1/N_{\rm cl}}$ and  $\sigma_{\rm map} \sqrt{3/N_{\rm cl}}$ respectively. Equation  \ref{eq:sigma_fil} can be applied to {\it any} filtering scheme and is the key to estimating the uncertainties of the eventual measurement using the KA-B methodology. The variance of the filtered maps, $\sigma_{\rm map}^2$, contains two terms adding in quadrature such that $\sigma_{\rm map}^2 = \sigma_1^2+\sigma_2^2$ with 1) $\sigma_1$ from the residual primary CMB anisotropies and 2) $\sigma_2$ from the instrument noise. The first of these gives rise to uncertainty which integrates down as $1/\sqrt{N_{\rm cl}}$ independently of the integration time and the number of pixels (fixed by the aperture size), $N_{\rm pix}$, involved in the final measurement, while uncertainty due to the second term integrates down as $N_{\rm pix}^{-1/2} t_{\rm integration}^{-1/2}$. AKEKE demonstrate that for the KABKE filtering scheme the filtering removes primary CMB down to the cosmic variance limit and the contribution from primary CMB becomes $\sigma_1\simeq 15 \sqrt{3/N_{\rm cl}}\mu$K. We use the WMAP data pixelized with $\simeq 7^\prime$ pixels, so the number of pixels subtended by a given aperture of radius $\theta_A$ is $N_{\rm pix} = \pi \theta_A^2/(7^\prime)^2 N_{\rm cl} \simeq 58  N_{\rm cl} (\theta_A/30^\prime)^2$. We will use below the four ($N_{\rm DA}=4$) W-band DA's which have the best angular resolution with the instrument noise per pixel of about $\sigma_W\simeq 130 \mu$K after 7-yr integrations (Jarosik et al 2011). Thus for the combined 7-yr W-band CMB data one obtains $\sigma_2 \simeq 9 (30^\prime/\theta_A) (4/N_{\rm A})^{1/2}\sqrt{3/N_{\rm cl}}\mu$K. When added in quadrature to $\sigma_1$ this term - for 7-year WMAP data and the final cluster apertures - gives a negligible contribution ($\lsim 15\%$) to the overall error budget.

We note that filtering cannot produce a dipole associated exclusively with clusters and whose magnitude further appears at zero monopole and increases with the cluster luminosity threshold. However, inappropriate filtering can decrease the S/N of the measured dipole rendering the measurement impossible. In this context it is worth pointing that Osborne et al use two filters designed to isolate and remove radio sources. Their Fig. 12 shows that their best filter does not recover bulk velocities with amplitude $\la 6,000$ km/s, while their other filter requires velocities of 30,000 km/sec or higher to be useful. However, massive Coma-like clusters moving at such speeds would generate KSZ anisotropies of $\delta T \ga 200 \mu$K that would be detectable in the {\it unfiltered} maps, while in the maps filtered with the Osborne et al adopted filters they are not. This by itself questions the suitability of the adopted filtering schemes in the KA-B method.

Our filter was designed to remove the primary CMB component, the main
contaminant on any KSZ measurement (see K-AB), and it does it down to cosmic variance (AKEKE, Fig. 1). Filters suited to detect point sources or to remove the TSZ component could
remove the CMB on large scales but boost it on small scales, reducing the
S/N measurement of the KSZ dipole at cluster locations. In any case,
the formalism described in AKEKE and above (eq. \ref{eq:sigma_fil}) enables to determine the efficiency of any filter. In this sense our filter is close to optimal because the primary CMB is removed down to the fundamental limit imposed by cosmic variance.

\subsection{A publicly available cluster sample}

To facilitate independent tests of the intermediate results of our data processing pipeline as well as of our final results concerning the presence and properties of the Dark Flow, we here provide step-by-step instructions on how to compile a basic version of the cluster catalogue used by KABKE. The resulting cluster sample should be nearly identical to the ones used by Keisler (2009) and Osborne et al.\ (2010) in their independent analyses of WMAP data.

The three publicly available catalogues of X-ray selected clusters compiled from ROSAT All-Sky Survey data (RASS, Voges et al.\ 1999) are the extended BCS sample (Ebeling et al.\ 1998, 2000) in the northern equatorial hemisphere, the REFLEX sample (B\"ohringer et al.\ 2004) in the southern equatorial hemisphere, and the CIZA sample (Ebeling, Mullis \& Tully 2002; Kocevski et al.\ 2007) in the regions of low Galactic latitude ($|b|<20^\circ$) excluded from both of the first two samples. All data contained in these catalogues can be obtained in electronic form at

\noindent
{\small \verb# http://vizier.cfa.harvard.edu/viz-bin/VizieR?-source=J/MNRAS/301/881/#}  (206 clusters)\\
{\small \verb# http://vizier.cfa.harvard.edu/viz-bin/VizieR?-source=J/MNRAS/318/333/#}  (99 clusters)\\
{\small \verb# http://vizier.cfa.harvard.edu/viz-bin/VizieR?-source=J/A+A/425/367/#} (447 clusters)\\
{\small \verb# http://vizier.cfa.harvard.edu/viz-bin/VizieR?-source=J/ApJ/580/774/#} (73 clusters)\\
{\small \verb# http://vizier.cfa.harvard.edu/viz-bin/VizieR?-source=J/ApJ/662/224#} (57 clusters)

Since all three cluster surveys used the same cosmology (Einstein-deSitter, $H_0 = 50$ km s$^{-1}$ Mpc$^{-1}$), the listed data sets can be immediately merged\footnote{As for equatorial cluster coordinates, care has to be taken to use the same epoch throughout, either B1950 or J2000.},  yielding a combined all-sky catalogue of 882 clusters. Removal of duplicates, caused by overlap between the REFLEX and eBCS catalogues at $0^\circ < \delta < 2.5^\circ$ leaves a sample of 771 unique clusters outside the KP0 CMB mask. We note that, while this sample is adequate to test the Dark Flow results, it is inferior to the one used by us in several respects:

\begin{description}
\item[Homogeneity and completeness:]  The X-ray flux limits of the eBCS, REFLEX, and CIZA samples differ significantly, as does the completeness as a function of redshift of the three surveys. The resulting systematic inhomogeneities of the combined sample are amplified by the fact that the three surveys employ different algorithms to compute total cluster fluxes (and hence also luminosities). By contrast, KABKE (and all subsequent studies by our team) use a homogenized catalogue created by applying a global flux limit to cluster fluxes recomputed from the RASS raw data (see KABKE for details).
\item[Contamination:] The published catalogues contain entries that have since been identified as erroneous. For instance, 99\% of the X-ray flux of the REFLEX cluster RXC\, J0334.9--5342 are contributed by an AGN, as revealed in a pointed X-ray observation with the Chandra Observatory. The inclusion of objects that are X-ray bright, but not galaxy clusters and thus not subject to the SZ effect, increases the noise in a bulk-flow measurement based on the KA-B method.
\item[Redshift accuracy:]  The published catalogues contain erroneous cluster redshifts. For instance, the REFLEX cluster RXC\,J0358.8--2955, listed as being at $z{=}0.168$ by B\"ohringer et al.\ (2004), was found to be at $z{=}0.425$ in the MACS survey (Ebeling et al.\ 2010). Redshift errors of this magnitude have a dramatic impact on the derived cluster X-ray luminosities which, as shown by KAEEK, correlate strongly with the monopole of the CMB signal at the cluster locations and can be used efficiently to isolate the most massive clusters that contribute most strongly to the KSZ dipole.
\end{description}

\begin{figure}
\plotone{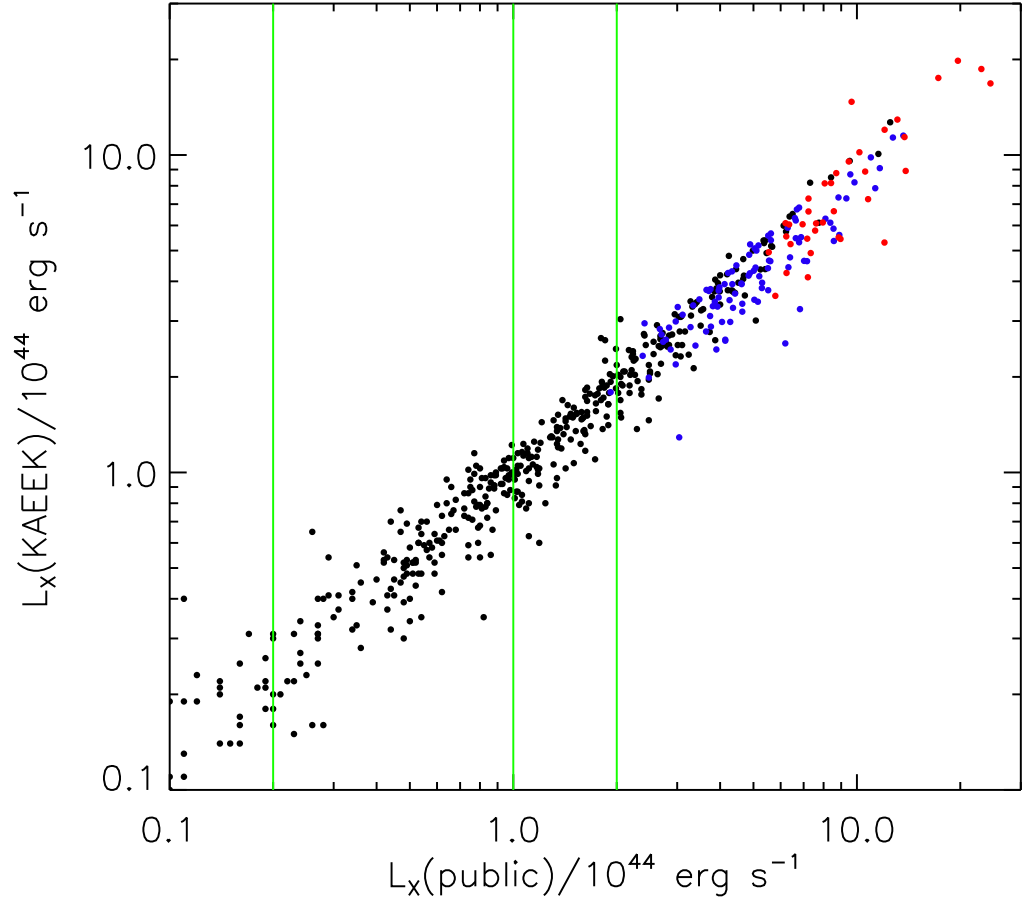} \caption{ \small{KAEEK vs publicly available X-ray luminosities for the broad
ROSAT band. Green vertical lines correspond to the $L_X$ bins used
in dipole computations. Black dots correspond to clusters at
$z\leq 0.16$, blue dots to $0.16 < z \leq 0.25$ and red dots to
$z>0.25$. At $z=0.25$ the radius subtended by the W-channel WMAP
beam (13$^{\prime\prime}$ radius) corresponds to 3 Mpc.  }} \label{fig:f2}
\end{figure}

Fig. \ref{fig:f2} illustrates the differences in cluster X-ray luminosity (scaled to the concordance $\Lambda$CDM model) between the KABKE sample and a simple cluster catalogue compiled from literature sources  as described above. While the impact of the corrections applied by KABKE (and KAEEK) are obvious, the good overall agreement supports our notion that the existence of a statistically significant bulk flow can be successfully tested from immediately available public cluster data.

Going beyond the extended homogeneous cluster sample used by KABKE, we have since launched SCOUT (Sunyaev-Zeldovich Cluster Observations as probes of the Universe's Tilt), a project designed to obtain an improved measurement and characterisation of the cluster bulk flow. SCOUT will use almost 1,500 clusters out to, and possibly beyond, $z{=}0.7$ to probe large-scale bulk motions with the KA-B methodology to yet larger distances, and with greater statistical accuracy, than KAEEK. Upon completion, this cluster catalogue too will be released to the community. In the meantime, we demonstrate here that the basic Dark Flow results can be obtained with publicly available cluster data (compiled as detailed above) and the filtering schemes described above.

\section{CMB dipole results using publicly available X-ray cluster data}

KAEEK and, particularly, AKEKE (also Keisler)
demonstrate that, for WMAP5 data and the KABKE filtering scheme, the
measurement errors are dominated by residual primary CMB fluctuations due to cosmic
variance. Adding channels thus does not appreciably increase the S/N of the
measurement.

\begin{figure}
\plotone{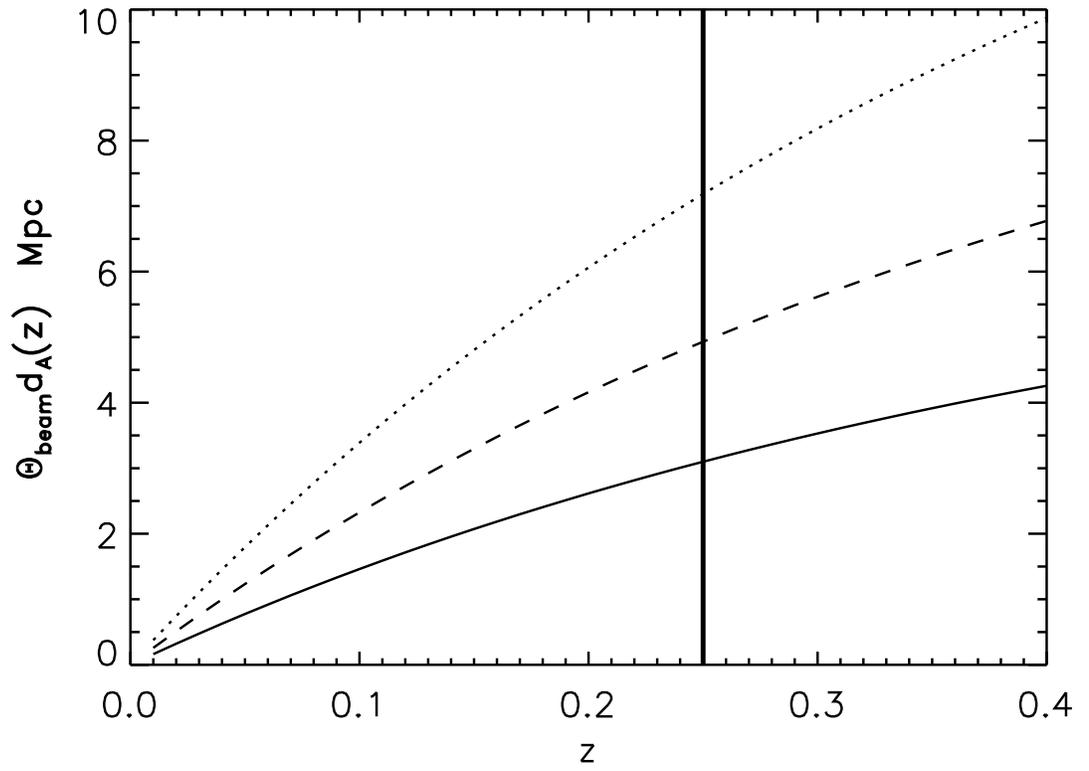} \caption{ \small{ Distance subtended at redshift $z$ by
the instrument beam of WMAP in the Q, V, W channels (dotted, dashed, solid). Thick vertical line marks the redshift limit of KAEEK and of this study.}} \label{fig:f3}
\end{figure}

Fig. \ref{fig:f3} shows the scales subtended by the beam of the three WMAP channels of the highest frequency. The WMAP W band has the best angular resolution
(13$^\prime$ radius beam), whereas all clusters are practically
unresolved in the Q band. Our analysis here thus uses only the four W channels of the WMAP differential assemblies (DA). At high redshift, however,  clusters are unresolved even in the W band; we thus follow KAEEK and impose a redshift limit of $z\le 0.25$.  Again following KAEEK we also remove clusters with $L_X <
2\times 10^{43}$ erg s$^{-1}$ (0.1--2.4 keV) because of the (relatively) more significant contamination
from X-ray emission by AGN.

\begin{figure}[h]
\plotone{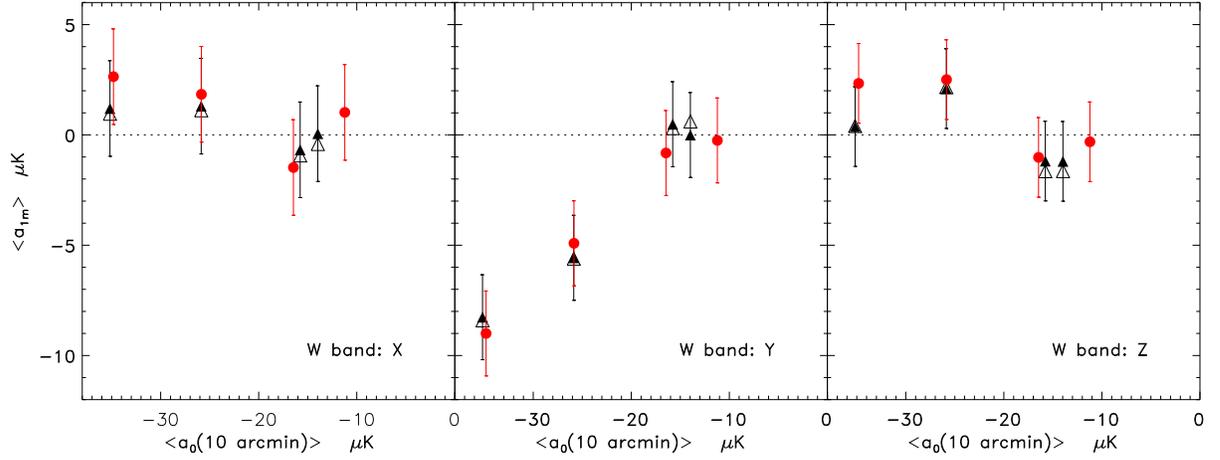}
\caption{\small{The values of the X,Y,Z dipole evaluated at
constant aperture of zero monopole for each sample marked with
green lines in Fig. \ref{fig:f2}. The horizontal axis shows the
value of the central (10$^{\prime}$ radius aperture)
monopole in the unfiltered W-band maps which is a reflection of
the $L_X$-threshold imposed. The last two
points at the largest negative $a_0$ correspond to $z\leq 0.16$
(142 clusters) and $z\leq 0.25$ (281 clusters). Red circles correspond to 5-yr WMAP maps. Black triangles are for 7-yr WMAP data: filled are for monopole and dipole subtracted maps; open triangles correspond to maps where quadrupole outside the mask was subtracted as well prior to filtering.}}
\label{fig:f4}
\end{figure}

We then bin the resulting cluster sample as shown by the green lines
in Fig. \ref{fig:f2} and evaluate the dipole {\it at the constant
aperture corresponding to zero monopole} for each subsample. The
results are computed for each W DA and averaged; errors are
computed as discussed in KAEEK and AKEKE (accounting for residual
primary CMB correlations). They are shown in Fig. \ref{fig:f4}
where, for each subsample, we plot the final dipole against the central monopole in
the unfiltered maps.

The results are clearly statistically significant and fully
consistent with those of KAEEK. In addition, there is a clear correlation
with the $L_X$ threshold, as expected if the signal is caused by the KSZ effect (the dipole is
computed at zero monopole; hence the TSZ contribution is small).
For clusters with $L_X\geq 2\times 10^{44}$ erg s$^{-1}$, the value of the $y$ component of the dipole obtained with this catalog in the W band is for 7-yr[5-yr] WMAP W-channel data:
\begin{eqnarray}
a_{1y}=-(8.3[9.0] \pm 2.6) \;\mu K \; ; \;  z\leq 0.16 \;; \; z_{\rm mean/median}=0.115/0.125 \; ; \; (l_0,b_0)=(278\pm 18, 2.5\pm 15)^\circ\\
a_{1y}=-(5.6[4.9] \pm 1.6) \;\mu K \; ; \;  z\leq 0.25 \;; \; z_{\rm mean/median}=0.169/0.176 \; ; \; (l_0,b_0)=(283\pm 19, 20\pm 15)^\circ\nonumber
\label{eq:dipole}
\end{eqnarray}
Here $(l_0, b_0)$ is the direction of the dipole in Galactic coordinates, and the results on the $y$-component represent 3- to 4-sigma detections using 142 and 281 clusters,
respectively. The errors are evaluated from eqs 4,6 of AKEKE. The decrease in amplitude between $z\leq 0.16$ and
$z\leq 0.25$ is consistent with the effects of beam
dilution decreasing the optical depth of the more distant
clusters. For comparison, for the same configuration KAEEK (Table 1) obtain, using 5-yr WMAP data and the first version of the SCOUT catalog, the $y$-component and the direction as $a_{1y}=(-8.0\pm 2.4)\;[(-4.1\pm 1.5)] \mu$K and $(l_0,b_0)=(292\pm 21, 27 \pm 15)^\circ\;[(296\pm 29, 39 \pm 15)^\circ]$ for $z\leq 0.16\;[0.25]$.

Comparison with Table 1 of KAEEK shows that the publicly available cluster sample used here is adequate to verify the basic dark flow result. However, the same comparison also demonstrates a clear superiority already of the preliminary SCOUT cluster sample (used in KAEEK) for a more accurate measurement of the components and properties of the dark flow.


\section{Calibration issues}

As discussed in KABKE2 and KAEEK, our current calibration
of the conversion from dipole amplitude to flow velocity may {\it
overestimate} the velocity of the flow. A more robust conversion will be provided by SCOUT where we will adopt an NFW profile to compute cluster properties, instead of the currently used isothermal $\beta$ model. Also, so far we have measured only the axis of motion but not the direction
of the flow along this axis because of the effects of filtering on the intrinsic KSZ terms (KAEEK). We discuss this last
point in more detail in this section, using the superior cluster catalog constructed for KAEEK.

A change of sign in the KSZ term can occur because we measure the dipole from the filtered maps, and the convolution of the intrinsic KSZ signal with a filter with wide side-lobes (as in KABKE) {\it can change the sign} of the KSZ signal for NFW clusters. The TSZ signal, which is more concentrated towards the inner cluster regions, will be less susceptible to this effect. Fig. \ref{fig:f5} illustrates this issue. The maps we use include SZ clusters and are convolved with the filter $F_\ell$ in $(\ell,m)$ space. This is equivalent to a convolution in the 2-D angular space $(\theta,\phi)$. After this convolution, the cluster properties clearly depend on the intrinsic profile of the clusters. As shown by AKKE, the latter are well described by an NFW model and are poorly matched by an isothermal $\beta$ model. Convolution will thus lead to different behavior of the SZ profiles, including the sign of the convolved SZ terms. In a measurement of the SZ signal from filtered maps, the intrinsic properties of clusters are first convolved with the beam ($B_\ell$ in $\ell$-space) and then with the filter. Fig. \ref{fig:f5} shows this filtering function, ${\cal F}(\theta) \equiv \sum_\ell (2\ell+1) F_\ell B_\ell P_\ell(\cos\theta)/\sum _\ell (2\ell+1) B_\ell$, where $P_\ell$ are Legendre polynomials. Because the convolution is performed in two dimensions, ${\cal F}(\theta)$ is multiplied by $\theta$ in the figure. The obvious side-lobes can affect the sign of the KSZ term in the outer parts differently than the more concentrated (for NWF profiles) TSZ terms. Because of the particular form of the KABKE filter, the sign of the KSZ dipole measured from the inner parts ($<10^\prime-15^\prime$ in angular radius where ${\cal F}$ remains positive) would be the opposite of the one of the signal we measure from the final apertures ($\simeq 30^\prime$).

Do the data support this interpretation? The 7-yr WMAP data have sufficiently low instrument noise to probe the SZ signal in the inner parts of the (stacked) clusters. Using only data from the W channels which provide the best angular resolution, we have evaluated the dipole in increasing cluster-centric apertures. The results are shown in Fig. \ref{fig:f6} where we plot the monopole and the dipoles against the aperture size for each of the four W DAs. We use the same cluster catalog as KAEEK but limited to systems with $L_X\ge 2\times 10^{44}$ erg s$^{-1}$ which yield the highest S/N, as shown in Table 1 of KAEEK. Even though the TSZ contributions can not be subtracted until we have  recomputed fundamental cluster properties for an NFW model, the data, if noisy, indeed suggest a sign change as the aperture is increased. Note that the zero-crossing is consistent for the $Y, Z$ components where the measurement is statistically significant.

In this context we stress again (see also KABKE)  the importance of using the entire aperture containing the full extent of the X-ray emitting gas that gives rise to the SZ effect, if one is to measure a statistically significant signal.

While there is thus indeed evidence for a sign change in the KSZ signal which affects the direction of the KAEEK-measured dipole, we emphasize again that a definitive answer will have to await a more complete, expanded and recalibrated SCOUT catalog. The flow direction can, however, be determined from applications of the KA-B method to Planck data, taking advantage of Planck's angular resolution of 5$^\prime$ --- a good match to the inner parts of clusters out to the limit of the SCOUT catalog --- and of the mission's 217 GHz channel for which the TSZ component vanishes. Since Planck (as well as Chandra on the X-ray side) will resolve
SCOUT clusters all the way out to $z\sim 0.6$, modelling of cluster properties with NFW profiles will be possible specifically for the most X-ray luminous systems which contribute most strongly
to the dipole signal. This measurement was already proposed by us to the Planck mission and will be performed when the Planck data become public and the final SCOUT catalog is assembled.

Nevertheless, the tentative evidence of the sign flip from filtering, already discussed by us earlier, allows to constrain the direction along the axis of motion determined in KAEEK. With the sign tentatively measured in Fig. \ref{fig:f6} the bulk flow motion will be in the direction of the detected CMB dipole at cluster positions; it is given in Table 1 of KAEEK  and the present data make little difference to it. Within the errors this direction will then coincide with the direction of the flow from Watkins et al at smaller scales ($\lsim 100$ Mpc) suggesting a coherent flow from the sub-100 Mpc scales to those probed in KAEEK ($\sim 800$Mpc).

\section{Discussion}

This paper demonstrates - using public X-ray data - the existence of a statistically significant dipole associated exclusively with clusters. The dipole signal is highly statistically significant and remains at apertures containing zero monopole. Its amplitude further increases with the X-ray luminosity threshold of the cluster subsamples as it should if produced by the SZ terms. However, the fact that it arises at zero monopole precludes any significant TSZ contributions to the signal as discussed in KABKE2 and AKEKE. We believe that the only explanation of this measurement is a large-scale bulk flow. Any alternative explanation of the signal has so far not been suggested in the literature. Adopting the large-scale-flow interpretation of the measurement, the properties of the flow (amplitude, direction and variation with depth) are fully consistent with Fig. 2 and Table 1 of AKEKE.

\begin{figure}
\plotone{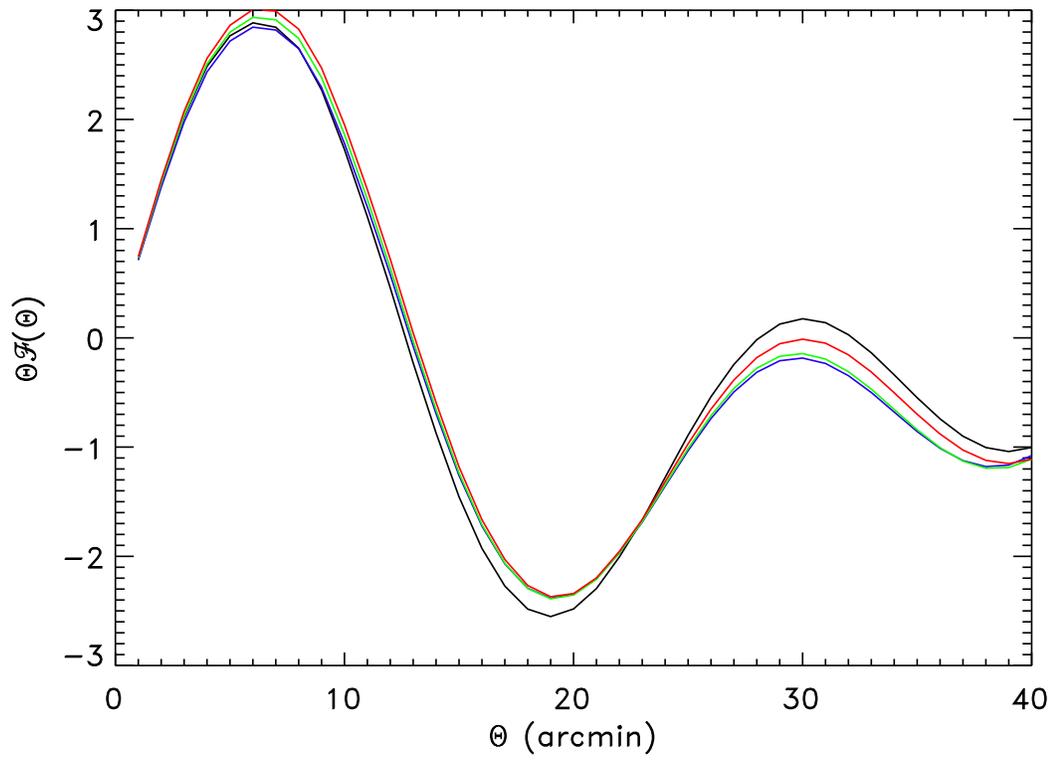} \caption{ \small{${\cal F}(\Theta)\equiv \sum (2\ell+1) F_\ell B_\ell P_\ell(\cos\Theta)/\sum (2\ell+1) B_\ell$. Black, blue, green, red correspond to the W1, W2, W3, W4 DA channels. respectively. }} \label{fig:f5}
\end{figure}

\begin{figure}
\plotone{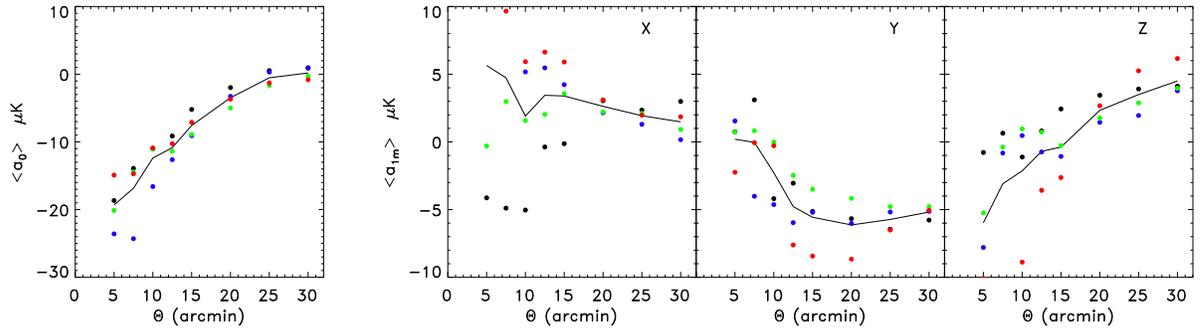} \caption{ \small{Monopole (left) and dipole components plotted vs the aperture radial size for the 7-year WMAP W band data. The numbers are evaluated at the
positions of $L_X\ge 2\times 10^{44}$ erg/sec and $z\leq 0.2$ clusters of the catalog used in KAEEK. Black, blue, green, red circles correspond W1, W2, W3, W4 DA channels respectively.}} \label{fig:f6}
\end{figure}

Adopting the bulk-flow interpretation of the measured dipole, with
the calibration coefficients for this configuration from Table 1 of
KAEEK, the flow amplitude would be $\sim$1,000 km s$^{-1}$ in the
direction given by Eq. \ref{eq:dipole}. The amplitude and the direction of the flow
are consistent with being constant at depths $zcH_0 \sim 300-550\,h^
{-1}$ Mpc. Note however the caveat that systematic calibration
uncertainties likely cause us to overestimate the amplitude by up to
30\% (KABKE2), and that interpreting the direction of the flow from the KSZ effect in the filtered maps remains subject to a sign
change for which we present first tentative empirical evidence in
Fig.\ref{fig:f5}.

In order to better probe and expand on our earlier ``dark flow" results we have designed an experiment named SCOUT (SZ Cluster Observations as probes of the Universe's Tilt). The SCOUT goals are to compile a sample of  $\sim 1,500$ X-ray selected galaxy clusters with spectroscopically measured redshifts out to significantly greater distances than the current $z{=}0.25$ limit, and to apply the KA-B method to the 9-year WMAP and Planck maps. The latter mission, with its low noise, higher angular resolution and wider frequency coverage, will be particularly useful in calibrating the measurements. First SCOUT results from a preliminary sample of $\sim 1,000$ clusters have been reported in KAEEK. While the SCOUT catalog is being assembled, we have shown in this paper that the basic dark flow results can be readily verified using publicly available cluster data. We make the sample generated from this database available upon publication at \url{http://www.kashlinsky.info/bulkflows/data\_public} and encourage the community to test our findings using the tools provided there.

In addition, this paper further addresses an important calibration issue resulting from  our filtering of the CMB maps. Using 7-year WMAP W-channel data, we show empirically that filtering may lead to a sign change in the KSZ term (see KAEEK). To resolve this issue and improve the calibration we need to decrease the noise in the measurement and properly recalibrate the catalog of cluster properties; both of these goals are achievable with a larger SCOUT sample. Application of our method to  Planck data in the 217 GHz channel, proposed by us earlier, will then allow accurate measurements of the velocity and direction of the flow.

We acknowledge NASA NNG04G089G/09-ADP09-0050 and
FIS2009-07238/GR-234/SyEC CSD 2007-00050 grants from Spanish Ministerio de Educaci\'on y
Ciencia/Junta de Castilla y Le\'on. We thank our collaborators on the SCOUT/"dark flow" project, Dale Kocevski and Alastair Edge, for their numerous valuable contributions to the project.

\clearpage


\clearpage

\clearpage

\clearpage

\clearpage

\clearpage

\end{document}